# Hybridization in van der Waals epitaxy of PtSe$_2$/h-BN and PtSe$_2$/graphene heterostructures

Meryem Bouaziz[1], Samir El Masaoudi [2], Aymen Mahmoudi[1], Eva Desgué[1,3], Marco Pala[1], Pavel Dudin[4], Mathieu G. Silly[4], Julien Chaste[1], Fabrice Oehler[1], Pierre Legagneux[3], Jose Avila[4], Iann C. Gerber[2], Abdelkarim Ouerghi[1]*

[1]Université Paris-Saclay, CNRS, Centre de Nanosciences et de Nanotechnologies, 91120, Palaiseau, France
[2]Université de Toulouse, INSA-CNRS-UPS, LPCNO, 135 Avenue de Rangueil, 31077 Toulouse, France
[3]Thales Research & Technology, 91767 Palaiseau, France
[4]Synchrotron SOLEIL, L'Orme des Merisiers, Départementale 128, 91190 Saint-Aubin, France

**Abstract**

Van der Waals (vdW) heterostructures, which combine bi-dimensional materials of different properties, enable a range of quantum phenomena. Here, we present a comparative study between the electronic properties of mono- and bi-layer of platinum diselenide (PtSe$_2$) grown on hexagonal boron nitride (h-BN) and graphene substrates using molecular beam epitaxy (MBE). Using angle-resolved photoemission spectroscopy (ARPES) and density functional theory (DFT), the electronic structure of PtSe$_2$/graphene and PtSe$_2$/h-BN vdW heterostructures are investigated in systematic manner. In contrast to PtSe$_2$/h-BN, the electronic structure of PtSe$_2$/graphene reveals the presence of interlayer hybridization between PtSe$_2$ and the graphene, which is evidenced by minigap openings in the π-band of graphene. Furthermore, our measurements show that the valence band maximum (VBM) of monolayer PtSe$_2$ is located at the Γ point with different binding energies of about -0.9 eV and -0.55 eV relative to the Fermi level on h-BN and graphene and substrates, respectively. Our results represent a significant advance in the understanding of electronic hybridization between TMDs and different substrates, and they reaffirm the crucial role of the substrate in any nanoelectronic applications based on van der Waals heterostructures.

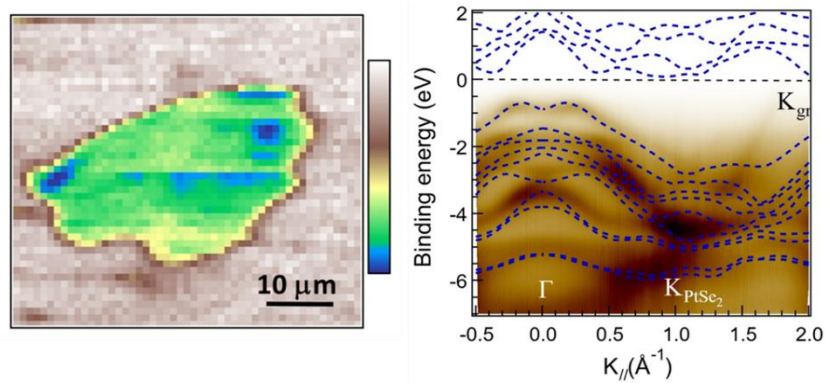

*Corresponding author: abdelkarim.ouerghi@c2n.upsaclay.fr

Transition metal dichalcogenide materials (TMDs) have attracted considerable attention in recent years due to their tunable band gaps, high carrier mobility and spin properties [1,2,3,4,5,6]. These materials represent promising candidates for the study of novel physical phenomena and functionalities in electronics, photonics, and superconductivity[7]. By combining individual monolayers of distinct materials into a van der Waals (vdW) heterostructure with atomically sharp interfaces it is possible to engineer the details of the band alignment [8,9,10,11]. This approach provides significant opportunity for the study of emerging electronic and optical properties at a fundamental level[12,7,13]

The first material of our vdW heterostructure is platinum diselenide PtSe, which is known for its thickness-dependent electronic properties, switching between semimetal and semiconductor. Density functional theory (DFT) calculations and scanning tunnelling spectroscopy (STS) measurements have revealed that $PtSe_2$ has an indirect bandgap of around 1.8 eV for 1 ML. This bandgap decreases to 0.8 eV for 2 ML, and is predicted to be zero for 4-7 ML [14,15,6]. To synthesize $PtSe_2$-based vdW heterostructures, few layers are typically obtained by mechanical exfoliation. However, experimental results are often compromised by defects produced during fabrication and the precise control over the rotational alignment of each stacked layers remains a significant challenge[13]. To avoid these limitations, the direct growth by Molecular Beam Epitaxy (MBE) appears as a viable alternative method for the growth of vdW heterostructures, with well-defined orientations and sharp interfaces[16,17,18,19,20]. The choice of epitaxial substrate is of major concern for MBE growth. The h-BN and graphene substrates are of particular interest as they are known to enable the direct growth of 2D/2D vdW heterostructures. Recently, the growth of $PtSe_2$ on graphene has been carried out successfully using MBE[21]. The electronic band structure of few layers of $PtSe_2$ have been measured, however the band structure of graphene substrate in the combined $PtSe_2$/graphene heterostructure has not been studied using ARPES techniques[22]. Recently Mallet et al.[23] have demonstrated the presence of hybridization between $PtSe_2$ and graphene substrate using STM/STS. On the other hand, different heterojunction based on $PtSe_2$ and 3D substrate such as ZnO have emerged[24], underlining the importance of substrate.

Concerning the direct growth of TMDs on h-BN substrate, there are a few studies on MBE growth[25–29]. Vergnaud et al. have studied the optical properties of $MoSe_2$ / h-BN structure, demonstrating the role of the temperature growth to improve the crystalline quality[21]. Lately, Luwiczak and coworkers[30] have studied the electronic properties of epitaxial $MoSe_2$ / h-BN. On the same substrate, Kim et al.[28] have successfully grown $PtSe_2$ on h-BN by MBE, with a focus on the optical emission properties. However, the electronic band structure has not been measured by ARPES so far due to the small size of the h-BN flakes.

Here, we investigate the electronic structure of MBE-grown $PtSe_2$/h-BN and $PtSe_2$/graphene heterostructures using nano-ARPES and DFT. These van der Waals interfaces, based on h-BN and graphene, enable the study of hybridization effects and charge transfer processes. The nanoscopic angle-resolved photoelectron spectroscopy (nano-ARPES) is a powerful technique allowing to focus the X-ray beam at 600 nm lateral resolution on the h-BN flakes to study the electronic band-structure of the $PtSe_2$/h-BN heterostructures[31,32]. Our results provide a precise study of the electronic properties for the two heterostructures grown in identical MBE conditions. Our ARPES/DFT demonstrate a strong electronic hybridization for single layer (1ML) and bilayer (2ML) $PtSe_2$ on graphene substrate.

**Results and discussions**

The PtSe$_2$ crystallizes in a 1T structure with $P\bar{3}m1$ space group (Figure 1(a)). The monolayer consists of a Pt atoms layer sandwiched between two Se layers with in-plane lattice constant of ~3.72 +- 0.02Å [33,34]. To understand the electronic properties and the hybridization changes between 1 ML PtSe$_2$/h-BN and 1, 2 ML PtSe$_2$/graphene heterostructures, we have performed DFT calculations. The theoretical in-plane lattice mismatch between PtSe$_2$ (a$_{PtSe2}$ = 3.72 Å) and the h-BN (a$_{h-BN}$=2.49 Å) or graphene (a$_{Graphene}$=2.47 Å) substrate is evaluated to ~50% respectively. The vertical heterostructure was modeled on the basis of a (2 × 2) PtSe$_2$ supercell on a (3 × 3) h-BN and graphene supercell, as shown in Figure 1(b). This leads to a reduced average lattice mismatch (~ +/- 0.40% respectively). Figure 1(c and d) shows the electronic band structure for 1ML PtSe$_2$/h-BN and 1ML PtSe$_2$/graphene, respectively. The direct comparison between the two calculated band structures of PtSe$_2$ on h-BN and graphene demonstrates that they are rather similar. However, the total density of state (DOS) of the two heterostructures, presented in Figure 1 (e and f), show some differences. In the DOS of the PtSe$_2$/h-BN system, four possible overlapping zones between the π-bands of h-BN and PtSe$_2$ can be identified. Also, in the DOS of 1ML PtSe$_2$/graphene system, we observe five possible overlapping peaks, indicated by dashed lines. This overlapping consists in the mixing of pz and px orbitals of C, or B, and N atoms with dz$^2$, dxz and dx$^2$-y$^2$ orbitals of Pt, mediated by p-orbitals of Se atoms. Interestingly, the hybridization between graphene and PtSe$_2$ is stronger than the interaction between h-BN and PtSe$_2$. Similar behavior is observed in the bilayer PtSe$_2$/graphene system, where overlapping peaks emerge at distinct positions, as shown in Figure S1.

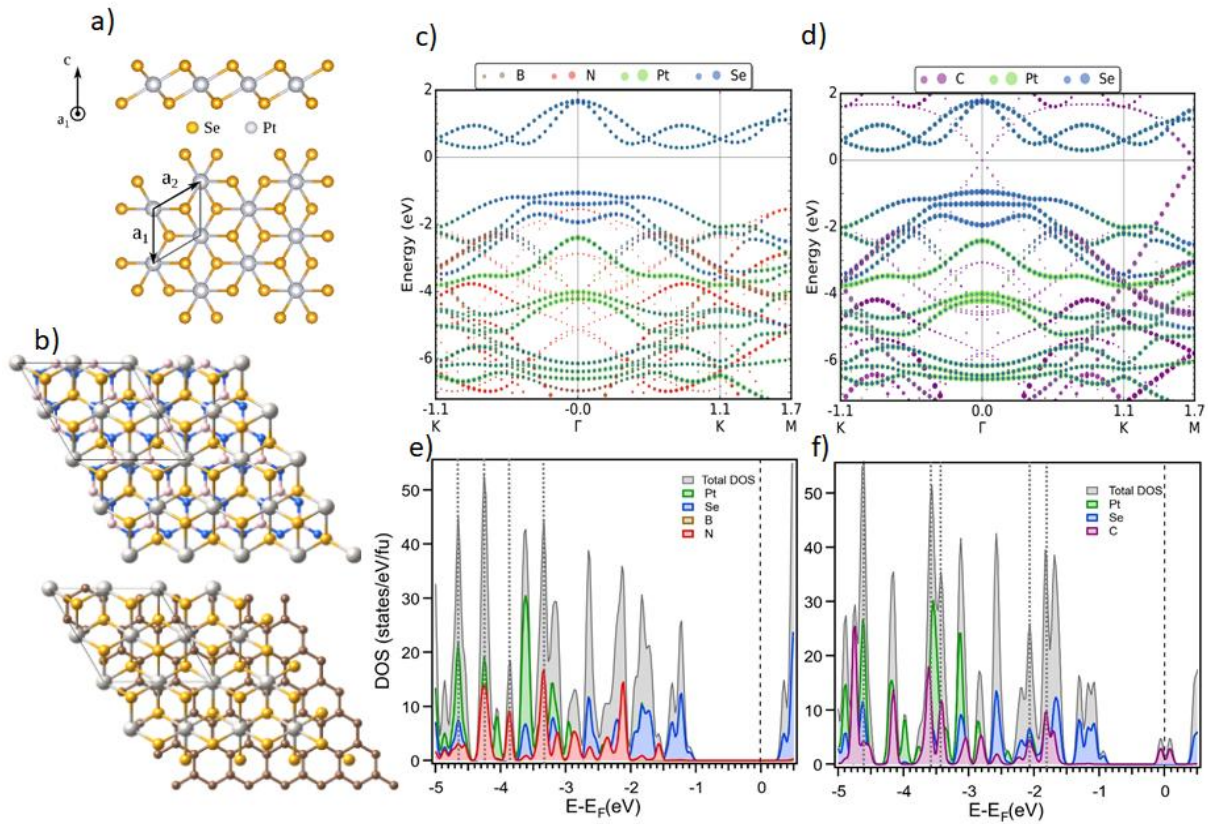

**Figure 1: Crystal structure and projected density of states of PtSe$_2$/h-BN and PtSe$_2$/ Graphene:** a) Crystal structure of single layer 1T PtSe$_2$, b) Schematic views of PtSe$_2$ grown on h-BN and graphene substrates, c and d) Band structures of PtSe$_2$/h-BN and PtSe$_2$/Graphene, respectively, with spin-orbit coupling following the high symmetry path KΓKM with each color corresponding to their respective atom contributions. e) Projected density

of states of 1ML PtSe$_2$/h-BN with spin-orbit coupling, from top to bottom: total DOS, Pt d orbitals, Se, N and B p orbitals DOS. f) Projected density of states of 1ML PtSe$_2$/Graphene with spin-orbit coupling, from top to bottom: total DOS, Pt d orbitals, Se and C p orbitals DOS,

To experimentally investigate these two heterostructures, the 1 and 2 ML PtSe$_2$ were directly grown on h-BN and graphene/SiC substrates during the same MBE growth (see details in Methods section). For this purpose, we use 2 cm$^2$ wide graphene/SiC(0001) square substrates. In order to reduce the charge transfer between the PtSe$_2$ and the substrate, the graphene/SiC(0001) were gentle hydrogenated[35]. A large number of h-BN flakes were exfoliated from the bulk material (10 flakes) on graphene. During the MBE growth, PtSe$_2$ grow over the whole surface, which include graphene and hBN/graphene areas, as shown in Figure 2(a). By optical contrast, areas of thick h-BN and thinner h-BN (yellow circle) are easily distinguished in Figure 2(b). Figure 2(c) shows reflection high-energy diffraction (RHEED) patterns taken after the growth of PtSe$_2$ on graphene. The presence of the sharp streaks suggests a well-order crystal of PtSe$_2$.

The Raman spectra of h-BN/graphene (Figure 2(d)) exhibit an E$_{2g}$ peak at ~1365 cm$^{-1}$ with a narrow linewidth of 7.34 ± 0.5 cm$^{-1}$, indicating a good structural quality for the h-BN material[36, 37,38]. Figure 2(e and f) show the Raman spectra of PtSe$_2$ on h-BN and graphene substrates under the excitation laser wavelength of 532 and 633 nm, respectively. In both Figures 2(e and f), three peaks are clearly identified at ~ 180, 207.5 and 236 cm$^{-1}$, which correspond to the E$_g$, A$_{1g}$ and longitudinal optical (LO) Raman active modes of PtSe$_2$, respectively[28]. The E$_g$ mode refers to (intra-layer) in-plane vibrations of the Pt-Se bonds while the A$_{1g}$ mode involves out-of-plane vibrations. The broader LO peak is a combination of two IR-active modes A$_{2u}$ and E$_u$. It can also be observed that the width of the E$_g$ and A$_{1g}$ peaks of PtSe$_2$ on h-BN (red spectra) are slightly narrower than that of PtSe$_2$ on graphene (black spectra): 2.5 cm$^{-1}$ compared to 3.6 cm$^{-1}$ for the E$_g$ peak. In contrast to the PtSe$_2$ on graphene, the LO peak of PtSe$_2$ on h-BN is well resolved and split into two peaks (at both excitations of 532 nm and 633 nm). These observations demonstrate an enhanced PtSe$_2$ crystal quality on h-BN surfaces compared to graphene/SiC.[24] For the Raman spectra on PtSe$_2$/h-BN, the measurements were made on two h-BN with different thicknesses: thin h-BN (blue spectrum) and thick h-BN (red spectrum). Based on the ratio between the two modes A$_{1g}$ and E$_{2g}$, it can be considered that the growth rate of PtSe$_2$ depends on the h-BN thickness: PtSe$_2$ monolayer is grown on the thinner h-BN, while bilayer of PtSe$_2$ is observed on the thicker h-BN[39]. These results are further discussed in the APRES section.

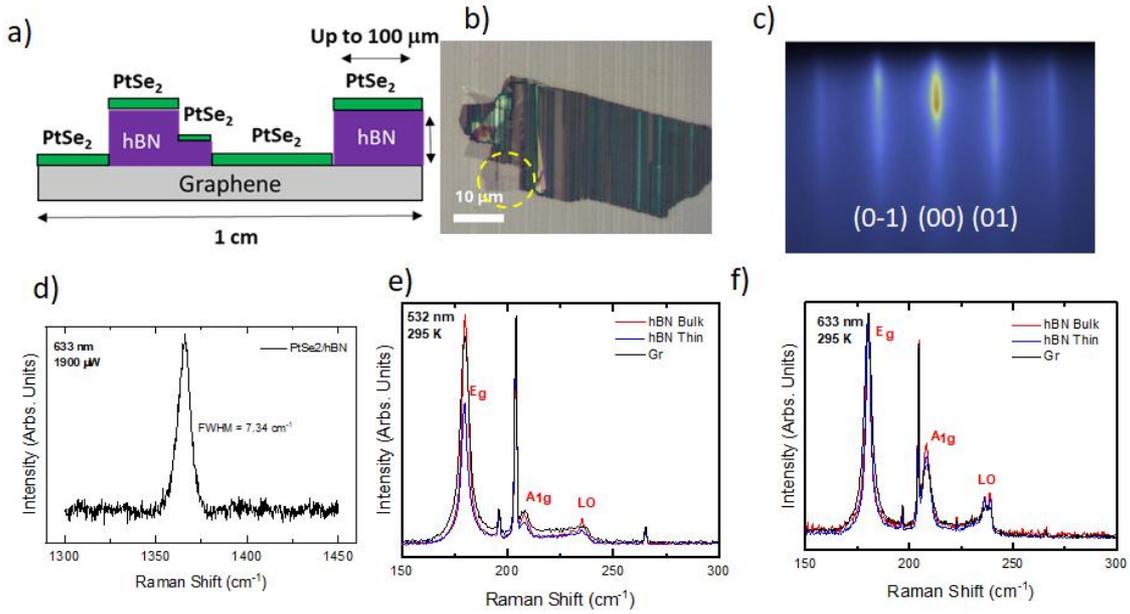

**Figure 2: Micro-Raman spectroscopy of PtSe$_2$/h-BN and PtSe$_2$/ graphene:** a) Scheme of PtSe$_2$ grown on graphene covered partially by h-BN flakes using MBE. The thickness of the h-BN flakes ranges from 1 to 10 nm. b) optical image of PtSe$_2$/h-BN showing different thickness of h-BN presented by different optical contrasts. c) RHEED patterns of PtSe$_2$/graphene substrate, d) micro-Raman spectrum taken on the h-BN/graphene layer. (e and f) micro-Raman spectra taken on the PtSe$_2$ on h-BN and graphene substrates for two laser excitation wavelengths 532 nm and 633 nm, respectively.

The nano-ARPES (T$_{sample}$ = 85 K)[40] is ideal to measure the intrinsic properties of PtSe$_2$/h-BN and PtSe$_2$/graphene. Figure 3(a) shows a spatial Se-3d intensity map of PtSe$_2$ grown on graphene and different flakes of h-BN that can be distinguished by optical contrast [12]. Figure 3(b) shows a zoomed XPS image of the spatial map (right) and the corresponding optical image (left), one can identify PtSe$_2$ on h-BN (green zone) and PtSe$_2$ on graphene (brown zone). The two regions display a unchanging intensity, attributed to the similarity of the electronic properties on the same h-BN flakes[17]. Figure 3(c) compares two typical Se 3d core level acquired on PtSe$_2$/graphene and PtSe$_2$/hBN. Two prominent peaks at 55.45 and 54.65 eV are attributed 3d$_{3/2}$ and 3d$_{5/2}$ orbitals of Se for PtSe$_2$ on h-BN, which is in good agreement with previous results[41]. In contrast a shift of Se 3d about 0.35 eV is observed between PtSe$_2$/h-BN and PtSe$_2$/graphene. Such difference is probably caused by the charge transfer or the intrinsic doping of the PtSe$_2$. These XPS measurements do not show any interdiffusion between PtSe$_2$ and both graphene or h-BN materials (i.e. only Pt-Se bonds are observed).

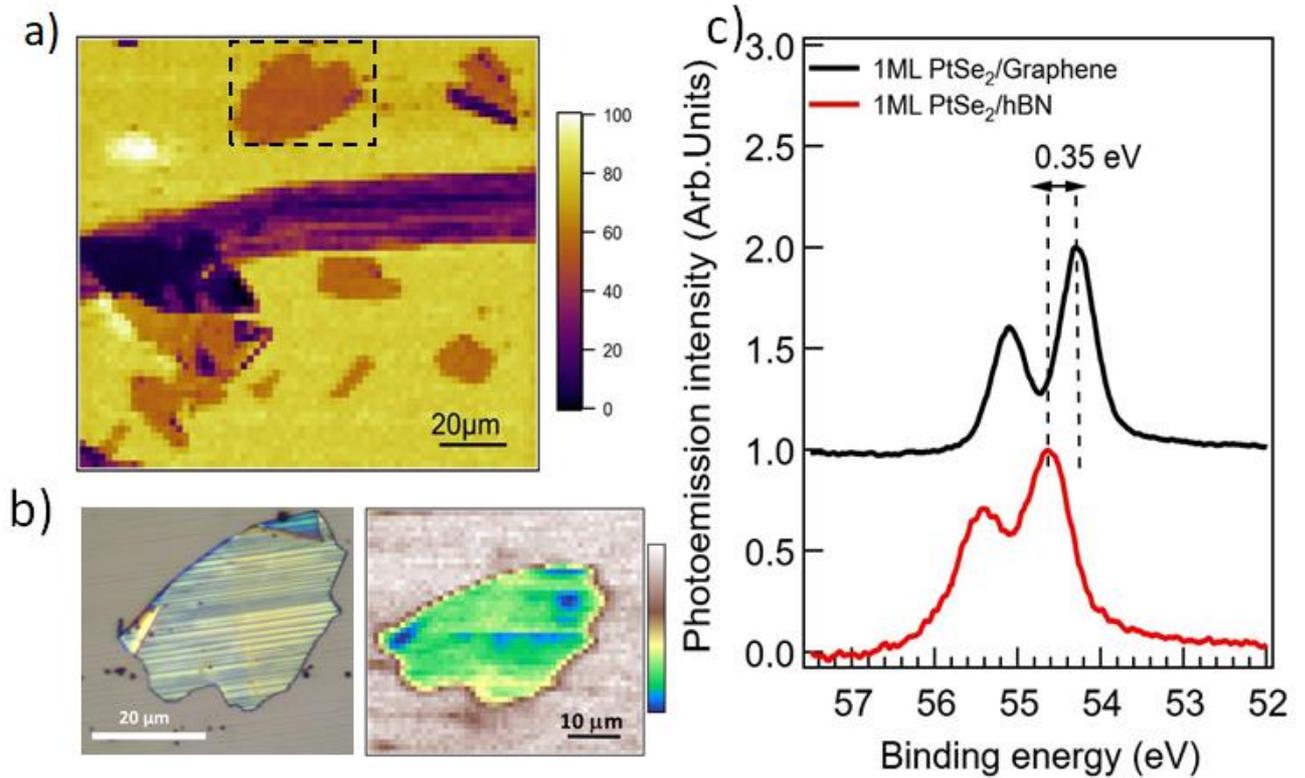

**Figure 3: Nano-ARPES image and spectra:** a) Spatially-resolved XPS map (of 100 × 100 μm² area) of the Se 3*d* core level obtained on the PtSe$_2$ sample by using a photon energy of 95 eV, showing PtSe$_2$ on different flakes of h-BN as exfoliated on graphene/SiC(0001) surface. b) Optical image and zoomed-in XPS spatial map of the h-BN zoomed flake from (a). c) Se 3*d* core level acquired on PtSe$_2$/h-BN and PtSe$_2$/graphene in the area shown in (**b**).

In Figure 4, we compare the band structure of PtSe$_2$/h-BN and PtSe$_2$/graphene along the Γ-K direction of PtSe$_2$. Using the ARPES constant energy maps recorded in the different zones of the sample, covered and uncovered by h-BN, we find that the BZ of PtSe$_2$ is aligned with the BZ of h-BN and graphene with no rotational variant (ie. Γ-K directions are superposed). Furthermore, for both heterostructure, we do not observe any replica due to a substrate-induced moiré-pattern. The corresponding electronic dispersions of PtSe$_2$ on both h-BN and graphene are collected along the high symmetry Γ-K direction and compared in Figure 4(a) and (b). The VBM, is located at - 0.9 eV and - 0.55 eV for 1ML PtSe$_2$/h-BN and 1ML PtSe$_2$/graphene, respectively. In both cases, the conduction band minimum (CBM) is above the Fermi energy (E$_F$).

The calculated band structures (blue dotted lines) in Figure 4 (c) and (d) show good quantitative agreement and all the main features are equally well reproduced. In monolayer, the VBM is dominated by the p$_x$ and p$_y$ orbitals of the Se atoms located at the Γ point and the CBM, originating from the d states of Pt and p states of Se, is located between the Γ and K points (figure 1c and d). Combining ARPES with DFT calculations, we confirm that monolayer PtSe$_2$ grown on h-BN has similar electronic properties of the free-standing PtSe$_2$. Theoretically, we can also confirm that the graphene layer remains semi-metallic: no gap is opened due to the hybridization with the

PtSe$_2$ layer at the Dirac point, since no sublattice symmetry breaking is induced, contrarily to h-BN/graphene heterostructures for instance[42,43].

For the width of the PtSe$_2$ bandgap, we refer to recent STS measurements which set the bandgap value for monolayer PtSe$_2$ at 1.80 eV[44]. Combined with our ARPES results, this places the conduction band minimum at ~0.90 and 1.25 eV respectively above the EF for the 1ML PtSe$_2$/h-BN and PtSe$_2$/graphene, respectively. This indicates that the PtSe$_2$, when grown on h-BN, exhibits a band alignment very close to the one expected for isolated intrinsic PtSe$_2$, with EF exactly situated at the middle of the bandgap. This shows that PtSe$_2$ grown by MBE should allow to fabricate PtSe$_2$ based devices exhibiting intrinsic properties. For PtSe$_2$/graphene, the energy difference between the two VBM is approximately +0.35 eV with respect to the "intrinsic" PtSe$_2$/h-BN indicating a p-type character. Different hypotheses can be formulated to explain the origin of this shift in the VBM: charge transfer, dielectric screening and intrinsic doping.

i) Raman modes frequencies are sensitive to charge transfer and strain. Here the E$_g$ and A$_{1g}$ modes remain unshifted in both structures, PtSe$_2$/h-BN and PtSe$_2$/graphene. This demonstrates the absence of charge transfer from hydrogenated graphene substrate to PtSe$_2$ within our experimental uncertainty. This is coherent with literature which quotes a reduction in charge transfer using such gently hydrogenated graphene.

ii) The observed VBM shift may also be attributed to dielectric screening, which directly changes the quasi-particle bandgap. Here, our DFT calculations (Figure 1c and d) show that the theoretical VBM shift between 1ML PtSe$_2$/h-BN and 1ML PtSe$_2$/graphene is only + 0.09 eV. This result is comparable but slightly lower than the reported dielectronic screening effect (+ 0.14 eV) between WS$_2$/h-BN and WS$_2$/graphite[45]. Still these findings suggest that the dielectric screening can only account for a small fraction of the observed VBM shift.

iii) From the Raman data, the widths of the E$_g$ and A$_{1g}$ peaks of PtSe$_2$ on h-BN are observed to be slightly narrower than those of PtSe$_2$ on graphene (black spectra): 2.5 cm$^{-1}$ FWHM compared to 3.6 cm$^{-1}$ for the Eg peak. These observations indicate a higher crystalline quality of PtSe$_2$ on h-BN compared to on graphene. This larger density of structural defect for the PtSe$_2$/graphene material could lead to an intrinsic doping of the PtSe$_2$/graphene and probably affects the position of the VBM.

Therefore, we conclude that dielectric screening and structural quality can account for the observed VBM shifts between PtSe$_2$/h-BN and PtSe$_2$/graphene, with PtSe$_2$/h-BN having nearly no offset to Fermi level and being close to reference intrinsic PtSe$_2$.

The following discussion will highlight the effect of PtSe$_2$ layer on the electronic band structure of h-BN and graphene. The π bands continuity in the h-BN band structure demonstrates that its electronic properties are fully preserved in the PtSe$_2$/h-BN heterostructure. In contrast, we observe that the π-bands of graphene have been slightly modified. In Figure 4(b), we remark the existence of minigap openings in the π-band of graphene substrate at higher binding energies. These minigaps demonstrate that the π-band of graphene at higher binding energies is altered by the interaction with the PtSe$_2$ layer.

In the two heterostructures, different unit cells may form a supercell, which behaves like a new unit cell in a hybrid crystal. The super-periodicity changes the overall band structure, opening up minigaps at high π-band of graphene[46]. In our system, the lattice parameters of h-BN (2.50 Å) and graphene (2.46 Å) are extremely close. Nevertheless, minigaps appeared only in the graphene band structure of PtSe$_2$/graphene and were absent in PtSe$_2$/h-

BN. Consequently, this observation confirms that the minigaps indeed occur in the system with the high electronic hybridization, that is, strong out-of-plane character (PtSe$_2$/graphene), while where the interaction is weak (PtSe$_2$/h-BN), no gaps in the h-BN π-band are induced.

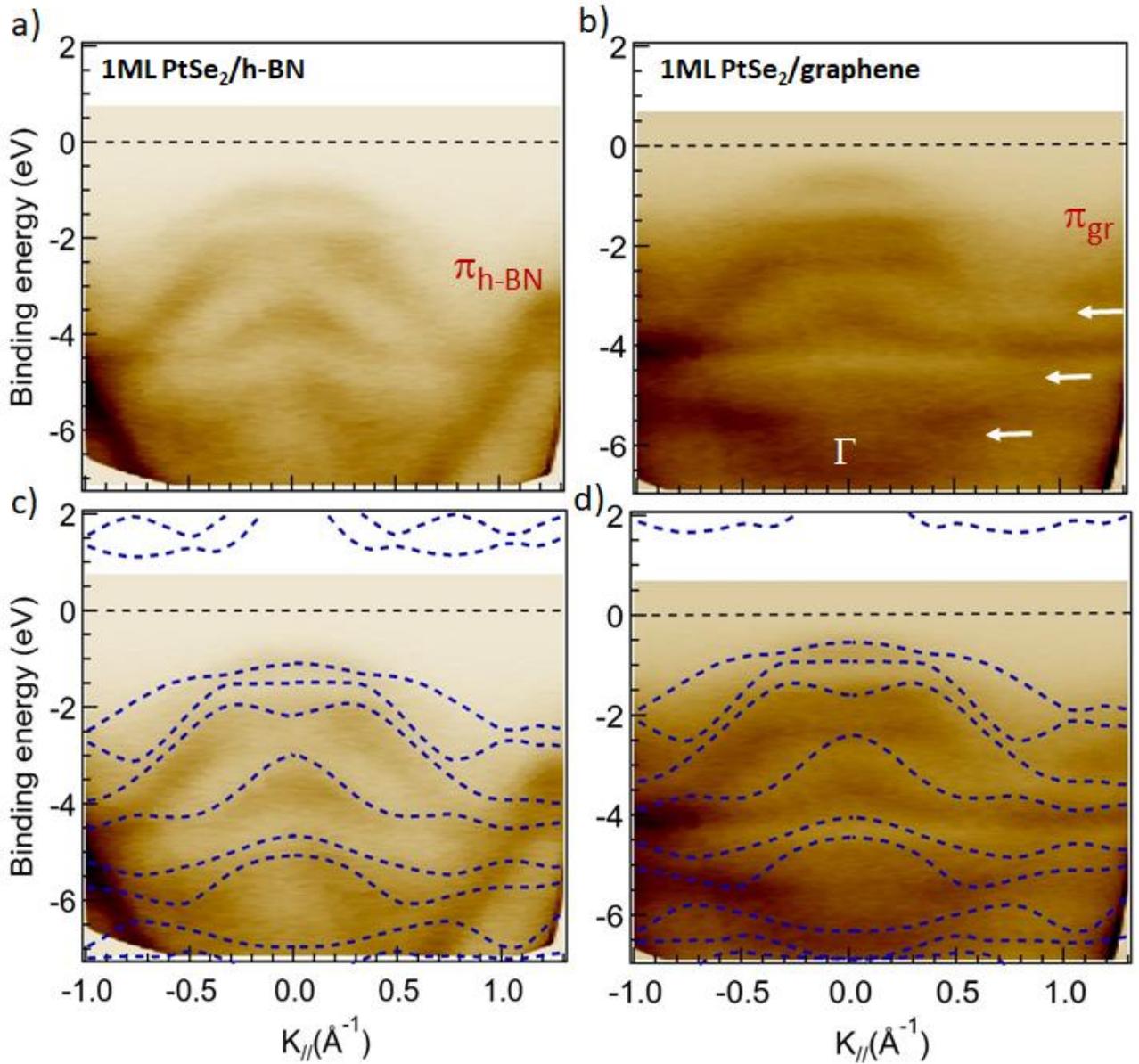

**Figure 4: ARPES image and valence bands of monolayer PtSe$_2$/h-BN and PtSe$_2$/Graphene:** (a) ARPES spectra obtained on the monolayer PtSe$_2$/ h-BN along the Γ-K direction. (b) The ARPES band structure of PtSe$_2$/graphene on the same direction (Γ-K), (c) and (d) Comparison between the ARPES and DFT electronic band structure (blue dashed lines) for PtSe$_2$/h-BN and PtSe$_2$/graphene, respectively.

To confirm this effect, further measurements of bilayer PtSe$_2$/h-BN and PtSe$_2$/graphene were made with nano-ARPES and ARPES, respectively. The Figure 5 (a and b) present nano-ARPES and ARPES of bilayer PtSe$_2$/h-BN and bilayer PtSe$_2$/graphene, showing that the VBM are located at 0.5 and 0.75 eV below the E$_F$, respectively.

The two-band structures show an additional band with a Mexican-hat shape. This band shift toward the $E_F$, indicating the tunable bandgap with varying film thickness confirmed using DFT calculation (see Figure S2). The comparison between the DFT/ARPES band structure is shown in Figure 5 (c and d). This ARPES/DFT are in a good agreement, confirming that we probed the 2ML structure on h-BN and graphene, respectively. Similarly, to the monolayer, no minigaps were observed for 2ML $PtSe_2$ on h-BN. In contrast, the ARPES of 2ML $PtSe_2$ on graphene shows minigap openings in the graphene's π- band, similarly to the 1ML $PtSe_2$ on graphene. Three distinct minigaps are observed along the Γ-K direction, labelled in sequential order in the range of 0.30, 0.35 and 0.40 ± 0.10 eV, respectively in Figure 5(b), and also found in the momentum distribution curves (MDC) in Figure S3. This further supports the argument for hybridization of the states between $PtSe_2$ and graphene.

According to our DFT calculations, the minigaps observed in the $PtSe_2$/graphene heterostructure arise at the intersection of specific $PtSe_2$ bands and the graphene π-band. These features are thus due to hybridization effects, which occur when $PtSe_2$ bands with out-of-plane orbital character overlap with the graphene π-band in terms of both energy and momentum (k-vector). This effect was observed by Batzill *et al.* in the case of the monolayer graphene on a $MoS_2$ substrate[46]. In our case, the DFT calculations confirm the possibility of hybridization processes between single or of bilayer $PtSe_2$ and the underlying graphene layer. Our result reminds the crucial role of the substrate in the fine tuning of the electronic properties in vdW heterostructures.

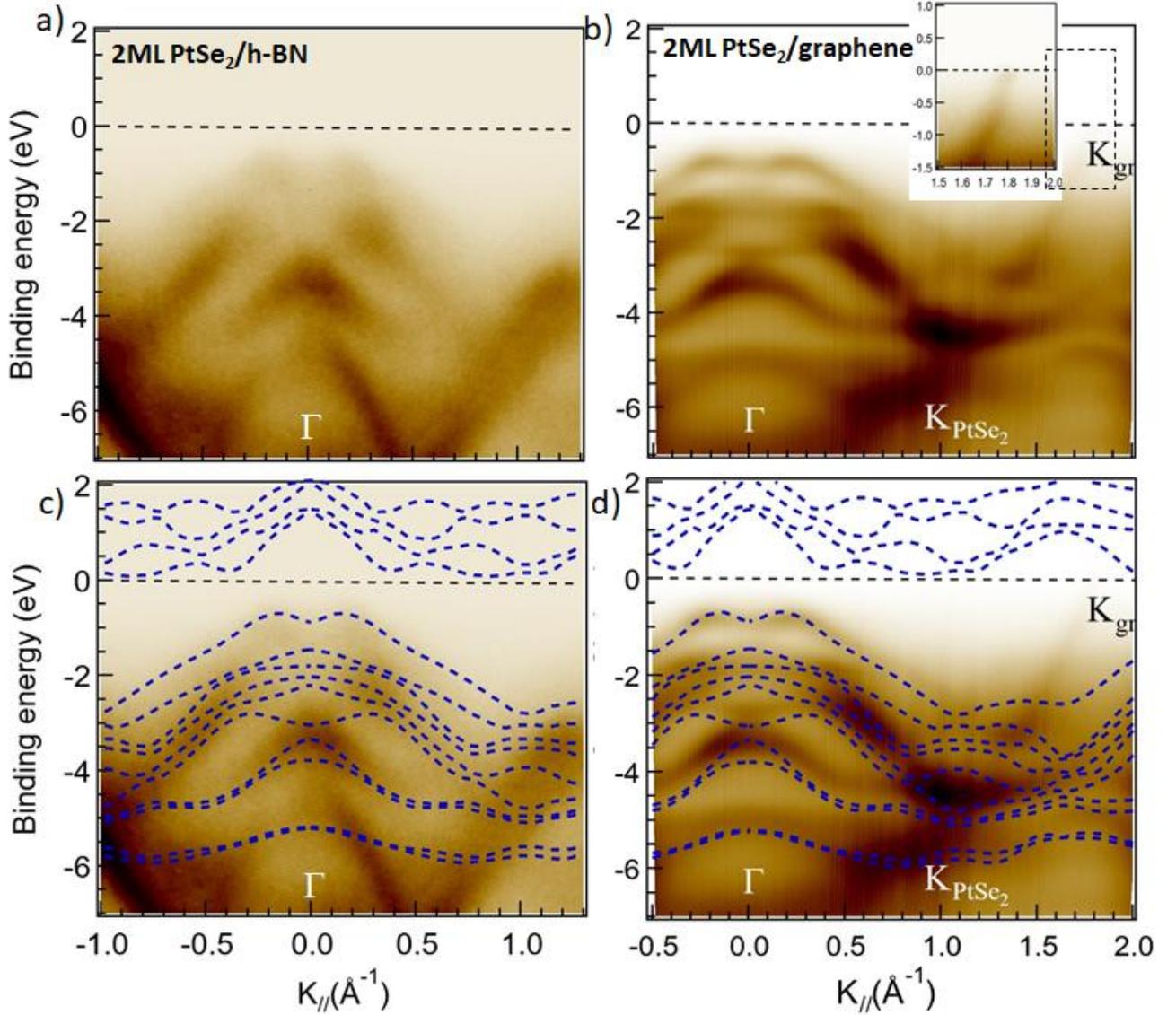

**Figure 5: Electronic structure of bilayer PtSe$_2$ on h-BN (a) and graphene (b)** obtained at hν = 90 eV, along Γ−K direction. Comparison of ARPES and DFT calculation of PtSe$_2$/h-BN and graphene along the Γ−K direction in (c) and (d), respectively. The black dashed lines indicate the EF position. The minigaps in b) are indicated by white arrows at 0.30, 0.35 and 0.40 ± 0.10 eV.

## Conclusions

In summary, we demonstrate the growth of PtSe$_2$/graphene and PtSe$_2$/h-BN heterostructures via van der Waals epitaxy. Monolayer and bilayer PtSe$_2$ grown on h-BN have analogous electronic properties as the free-standing mono- and bilayer PtSe$_2$, with Fermi level in the middle of the bandgap. In contrast to the PtSe$_2$/h-BN heterostructure, the band structure of PtSe$_2$/graphene reveals the presence of interlayer hybridization. While the Dirac cone of graphene remains intact, we demonstrate the creation of minigap openings in the π-band of graphene, due to electronic hybridization of PtSe$_2$ layer and graphene substrate. Thus, this family of materials provides an approach to tuning the electronic properties of vdW heterostructures by selecting the appropriate substrate.

**Materials and methods**

**Synthesis.** The PtSe$_2$ sample was synthesized in a 2-inch MBE reactor supplied by Dr Eberl MBE-Komponenten. The selenium flux was fixed at $\Phi(Se) = 0.5$ Å.s$^{-1}$ and was supplied by a valved selenium cracker source filled with ultra-high purity (7N) Se, heated to 290 °C (the cracker temperature was set at 600 °C). An electron beam evaporator generated the Pt flux from high purity (4N) Pt, which was set to $\Phi(Pt) = 0.003$ Å.s$^{-1}$, meaning that $\Phi(Se)/\Phi(Pt) \sim 170$. The 20 x 20 mm$^2$ graphene/SiC(0001) substrate supporting hBN flakes was outgassed under UHV conditions at 700 °C for 15 min. The growth was performed at 320 °C and began with 5 cycles, each including 2 steps: during 2 min, the substrate was exposed to both Se and Pt fluxes to grow PtSe$_2$, then the sample was annealed under Se flux for 5 min. These two steps were repeated five times, after which the growth continued with simultaneous Se and Pt fluxes to achieve the desired PtSe$_2$ thickness. After growth, the temperature was cooled down to 200 °C over 20 min under the same Se flux used during growth, to avoid Se desorption from the film.

**Experiments.**

Micro-Raman spectroscopy was performed using a commercial Horiba confocal microscope equipped with a 100× objective and laser excitations at 532 and 633 nm. The laser beam was focused to a ~1 μm diameter spot on the sample surface.

Nano-ARPES measurements on 1 ML PtSe$_2$/graphene and 1–2 ML PtSe$_2$/h-BN van der Waals heterostructures were conducted at the ANTARES beamline of the SOLEIL synchrotron facility, using linearly polarized light with a photon energy of 95 eV with a spot size of 600 nm. The measurements were carried out under a base pressure of $3 \times 10^{-10}$ mbar, with the sample maintained at 85 K.

ARPES measurements of bilayer PtSe$_2$/graphene were performed at the TEMPO beamline of the same facility, using an MBS analyzer. The incident photon beam was focused to a spot size below 80 μm on the sample surface.

**Computational details.**

First-principles calculations for the freestanding PtSe$_2$ monolayer shown in Figure 4 were performed with the density functional theory (DFT) using the Quantum ESPRESSO suite. A fully relativistic, norm-conserving pseudopotential was employed, along with non-collinear spin calculations to account for spin–orbit coupling, using the Perdew–Burke–Ernzerhof (PBE) exchange–correlation functional[47]. van der Waals interactions were included via the non-local vdW-DF3 functional[48]. Self-consistent calculations were carried out using a $10 \times 10 \times 10$ Monkhorst–Pack k-point grid and a plane-wave energy cutoff of 80 Ry. Both atomic positions and lattice parameters were relaxed until force and energy convergence thresholds of $10^{-3}$ and $10^{-8}$ a.u., respectively, were reached.

**Computational details.**

Periodic density functional theory (DFT) calculations for the heterostructures were performed using the plane-wave pseudopotential approach, as implemented in the Vienna Ab Initio Simulation Package (VASP). The Perdew–Burke–Ernzerhof (PBE) exchange–correlation functional within the generalized gradient approximation (GGA) was employed, and van der Waals interactions were accounted for using the many-body dispersion scheme

(MBD@rsSCS) developed by Tkatchenko et al.[48] Core–valence interactions were treated using the projector augmented-wave (PAW) method[49,50] and a plane-wave energy cut-off of 500 eV was applied.

The Pt pseudopotential included 18 valence electrons ($5s^2\ 5p^6\ 5d^{10}$), while standard valence configurations were used for the other elements. For bulk $PtSe_2$, a $21 \times 21 \times 11$ k-point mesh centred at the gamma point was used to determine the lattice parameters, while an $11 \times 11 \times 1$ grid was applied to the $(3 \times 3)/(2 \times 2)$ superstructures. The optimised lattice constants were a = 3.766 Å and c = 4.975 Å, which is in good agreement with experimental data[51]. A vacuum spacing of at least 18 Å was introduced along the z-axis to suppress interactions between periodic images. Spin–orbit coupling was included in the final band structure calculations, which were then unfolded using [52].

**ACKNOWLEDGMENTS**

This work was supported by the French National Research Agency (ANR) through the following projects: DEEP2D (ANR-22-CE09-0013), MixDferro (ANR-21-CE09-0029), FastNano (ANR-22-PEXD-0006) and EXODUS (ANR-23-QUAC-0004). It was also supported by the French national nanofabrication network, RENATECH. S. E. M. acknowledges additional support from the ANR project (ANR-21-CE09-0003) and the NanoX Graduate School programme (ANR-17-EURE-0009), both of which are part of the "Programme des Investissements d'Avenir". S. E. M. and I. C. G. gratefully acknowledge the computational resources provided by the CALMIP initiative (project P0812) and by CINES, IDRIS and TGCC, which were granted by GENCI through allocation 2024-A0140906649.

**Supporting Information:** Band structure of 2 ML $PtSe_2$/Graphene with spin-orbit coupling. Energy curve analysis (EDC) of the ARPES spectra as reported Figure 5(b). ARPES measurements and valence band structures of single-layer and bilayer $PtSe_2$ on h-BN and graphene substrates

**Competing financial interests:** There are no conflicts to declare.

# Hybridization in van der Waals epitaxy of PtSe$_2$/h-BN and PtSe$_2$/graphene heterostructures

Meryem Bouaziz[1], Samir El Masaoudi[2], Aymen Mahmoudi[1], Eva Desgué[1,3], Marco Pala[1], Pavel Dudin[4], Mathieu G. Silly[4], Julien Chaste[1], Fabrice Oehler[1], Pierre Legagneux[3], Jose Avila[4], Iann C. Gerber[2], Abdelkarim Ouerghi[1]*

[1]Université Paris-Saclay, CNRS, Centre de Nanosciences et de Nanotechnologies, 91120, Palaiseau, France
[2]Université de Toulouse, INSA-CNRS-UPS, LPCNO, 135 Avenue de Rangueil, 31077 Toulouse, France
[3]Thales Research & Technology, 91767 Palaiseau, France
[4]Synchrotron SOLEIL, L'Orme des Merisiers, Départementale 128, 91190 Saint-Aubin, France

*Corresponding author: abdelkarim.ouerghi@c2n.upsaclay.fr

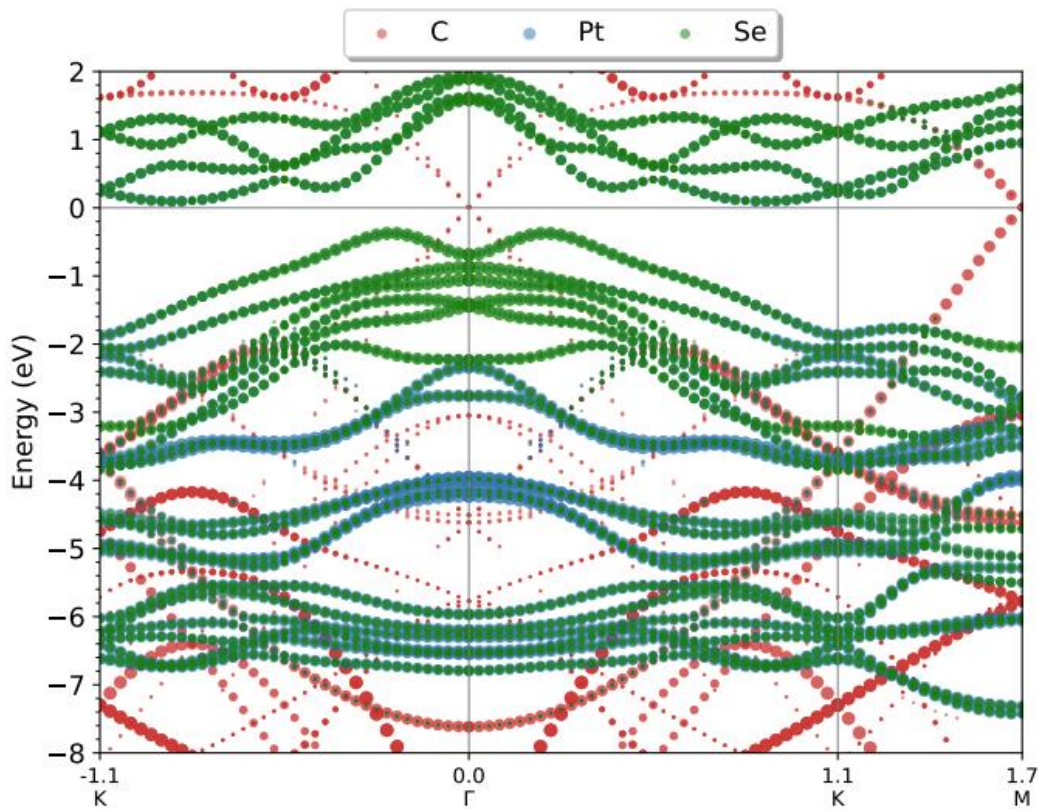

**Figure S 1:** a) Band structure of 2 ML PtSe$_2$/Graphene with spin-orbit coupling following the high symmetry path: K --> Gamma --> K --> M with each color corresponding to their respective atom contribution,

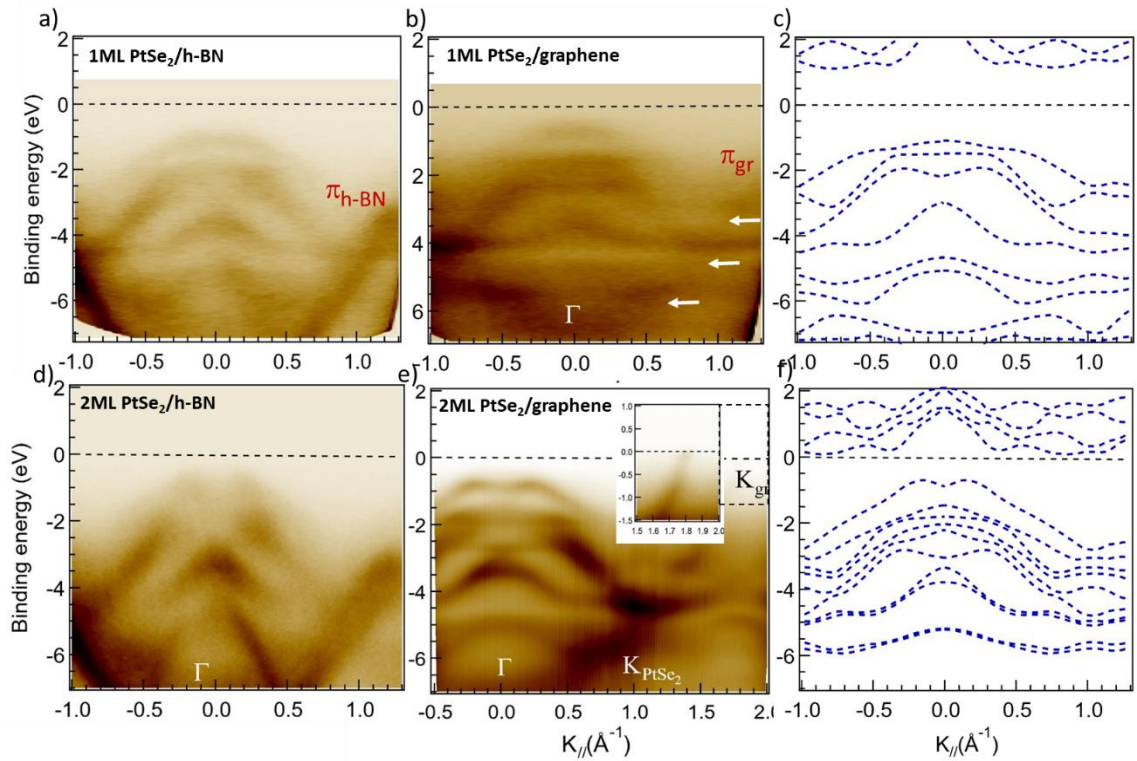

**Figure S2:** ARPES measurements and valence band structures of single-layer and bilayer PtSe$_2$ on h-BN and graphene substrates. (a) ARPES spectrum of monolayer PtSe$_2$/h-BN along the Γ–K direction of the hexagonal Brillouin zone. (b) Corresponding ARPES spectrum of monolayer PtSe$_2$/graphene along the same direction. (c) DFT-calculated band structure of monolayer PtSe$_2$ along Γ–K. (d, e) ARPES band structures of bilayer PtSe$_2$ grown on h-BN and graphene, respectively, along Γ–K. (f) DFT-calculated band structure of bilayer PtSe$_2$.

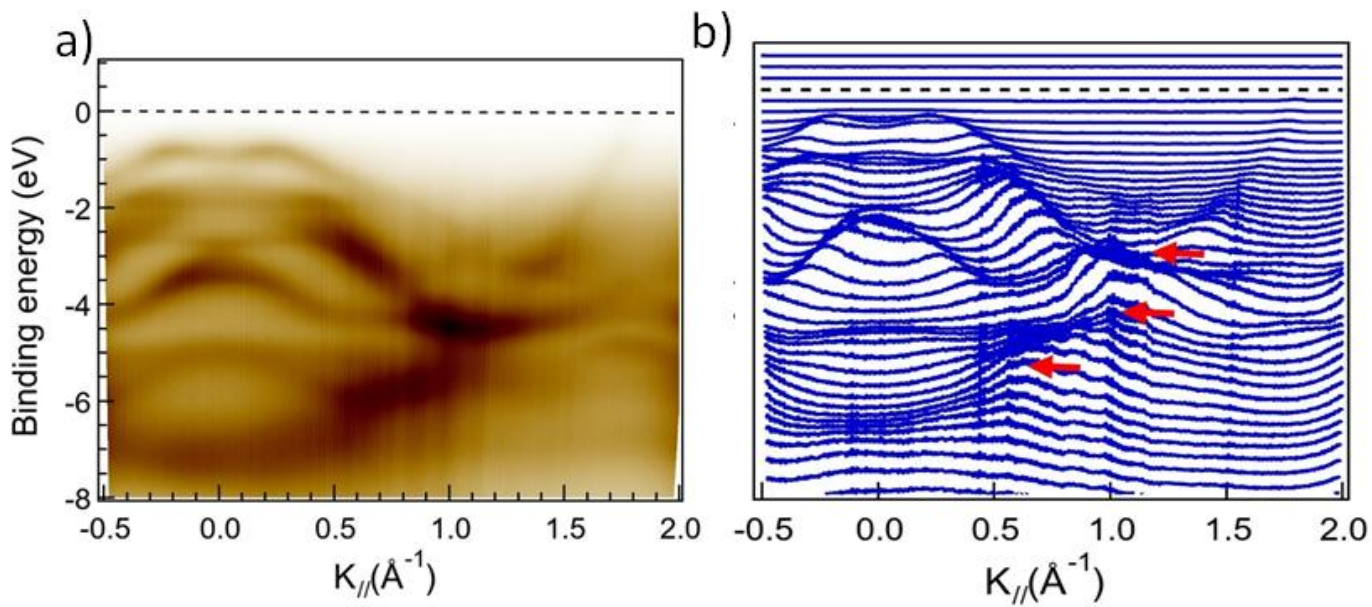

**Figure S3:** Electronic structure of bilayer PtSe$_2$/graphene: (a) ARPES spectra of h-BN at hν = 90 eV, (b) Stack plots of MDCs for the valence band extracted from (a).